\documentclass[aps,superscriptaddress,preprint]{revtex4}

\usepackage{graphicx}
\usepackage[latin1]{inputenc}
\usepackage{amsmath}

\newcommand{\bit}{\begin{itemize}}
\newcommand{\eit}{\end{itemize}}

\begin{document}

\title{Determining the underlying Fermi surface of strongly
correlated superconductors}

\author{Claudius Gros}

\affiliation{Institute for Theoretical Physics, J.W. Goethe University Frankfurt, 60438 Frankfurt, Germany}

\author{Bernhard Edegger}

\affiliation{Institute for Theoretical Physics, J.W. Goethe University Frankfurt, 60438 Frankfurt, Germany}

\affiliation{Department of Physics, Princeton University, Princeton, NJ 08544, USA}

\author{V. N. Muthukumar}

\affiliation{Department of Physics, City College of the City University of New York, New York, NY 10031, USA}

\author{P. W. Anderson}

\affiliation{Department of Physics, Princeton University, Princeton, NJ 08544, USA}

\email{}

\date{}

\maketitle

{\bf The notion of a Fermi surface (FS) is one of the most
ingenious concepts developed by solid state physicists during the
past century.  
It plays a central role in our
understanding of interacting electron systems. Extraordinary
efforts have been undertaken, both by experiment and by theory, to
reveal the FS of the high temperature superconductors (HTSC), the
most prominent strongly correlated superconductors.
Here, we discuss some of the prevalent methods
used to determine the FS and show that they lead generally to
erroneous results close to half filling and at low temperatures,
due to the large superconducting gap (pseudogap) below (above) the
superconducting transition temperature. Our findings provide a
perspective on the interplay between strong correlations and
superconductivity and highlight the importance of strong coupling
theories 
for the characterization as well
as the determination of the underlying FS in ARPES experiments. }

During the last decade, Angle Resolved Photoemission Spectroscopy
(ARPES) has emerged as a powerful tool
\cite{Damascelli03,campuzano_rev}
to study the electronic structure of the HTSC
\cite{nature_rev}. This is
because ARPES is a direct method to probe the FS, the locus in
momentum space where the one electron excitations are gapless
\cite{Pines_Nozieres}.
However, since the low temperature phase of the HTSC has a
superconducting or pseudogap with $d$-wave symmetry, an FS can be
defined only along the nodal directions or along the so-called
Fermi arcs
\cite{Damascelli03,campuzano_rev,Loeser96,Marshall96,Norman98}.
The full `underlying FS' emerges only when the pairing
interactions are turned off, either by a Gedanken experiment, or
by raising the temperature. Its experimental determination
presents a great challenge since ARPES is more accurate at lower
temperatures. Since the FS plays a key role in our understanding
of condensed matter, it is of importance to know what is exactly
measured by ARPES in a superconducting or in a pseudogap state.
The problem becomes even more acute in HTSC due to the presence of
strong correlation effects \cite{Anderson87,Lee06,Zhang88,elstr}.
Hence, it is desirable to examine a
reference $d$-wave superconducting state with aspects of strong
correlation built explicitly in its construction. Motivated by
these considerations, we study the FS of a strongly correlated
$d$-wave superconductor \cite{Zhang88,elstr} and discuss our
results in the context of ARPES in HTSC.

We begin by highlighting the differences between a Fermi and a
Luttinger surface. The FS is determined by the poles of the one
electron Green's function \cite{Pines_Nozieres}. The Luttinger
surface is defined as the locus of points in reciprocal space,
where the one particle Green's function changes sign
\cite{Dzyaloshinskii03}. In the Fermi liquid state of normal
metals, the Luttinger surface coincides with the FS. In a
Mott-Hubbard insulator the Green's function changes sign due to a
characteristic $1/\omega$-divergence of the single particle self
energy \cite{Gros94} at momenta $k$ of the non-interacting Fermi
surface. In the HTSC the gapped states destroy the FS but only
mask the Luttinger surface. Hence, it seems natural to relate the
Luttinger surface of the superconducting and of the pseudogap
states with the concept of an `underlying FS', and ask if such a
surface can be determined by ARPES.

To answer this question, we recall that the elementary
excitations in a superconductor are given by the dispersion
relation,
\begin{equation}
E_k\ =\ \sqrt{\xi_k^2+\Delta_k^2},\qquad
\xi_k=\epsilon_k-\mu~,
\label{E_k}
\end{equation}
where $\epsilon_k$ are the momentum dependent orbital energies of
electrons in the absence of a superconducting order parameter
$\Delta_k$; $\mu$ is the chemical potential. The corresponding
Luttinger surface is determined by the condition $\xi_k\equiv0$,
which is also the definition of the normal state FS when
$\Delta_k\equiv0$. In the following, we discuss two methods
commonly used to determine the underlying FS, {\it viz.}, the
Luttinger surface, of the HTSC by ARPES
\cite{Damascelli03,campuzano_rev,Mesot01}.

In the so-called `maximal intensity method' the intensity of
ARPES spectra at zero frequency is used to map out the underlying FS. It can be shown
that this quantity is,
\begin{equation}
\sim \frac{\Gamma_k}{E_k^2+\Gamma_k^2}~, \label{spInt}
\end{equation}
where, $\Gamma_k$ is determined both by the experimental resolution
and the width of the quasiparticle peak. When the momentum
dependence of $\Gamma_k$ is small compared to that of $E_k$ (as is
usually the case), the maximal intensity is given by the set of
momenta $\hbar k$ for which $E_k$ is minimal.

To examine the accuracy of this method in determining the
underlying FS, we calculate this quantity for a strongly
correlated $d$-wave superconducting state. All calculations are
done with model parameters for HTSC using the renormalized mean
field theory (RMFT) \cite{Zhang88,elstr}, for which the
quasiparticle dispersion $E_k$ retains the form of Eq.~\ref{E_k}.
In Fig.~1, we show our results for the spectral intensity at zero
frequency as well as the locus of the Luttinger surface. The
former is deduced from the inverse of $E_k$.

For large hole doping, $x=0.25$, the superconducting gap is small
and the Luttinger surface is close to the points in momentum space
for which the zero frequency intensity is maximal. But for smaller
doping, $x=0.05$, the gap is substantial and the Luttinger surface
deviates qualitatively from the maximal intensity surface due to
the momentum dependence of $\Delta_k$ (see ridges in Fig.~1). We
have verified that this behaviour persists for a wide range of
$|\Delta_k|$, and not just the values estimated from RMFT used in
Fig.~1. Although not widely discussed in the literature, this
splitting may be deduced from experimental data, \textit{e.g.},
the intensity plots in $E-k$ space along symmetric lines
$(0,0)\rightarrow(\pi,0)\rightarrow(\pi,\pi)$ in~\cite{yoshida05}.
It follows that when the gap or the pseudogap is large, the
criterion of maximal spectral intensity alone does not suffice to
identify the correct FS and it is necessary to supplement the
analysis of the zero frequency ARPES intensity (Eq.~2) with a
dispersion relation such as Eq.~1. These considerations explain
why the (outer) maximal intensity ridges seen in ARPES (at low
temperatures in the underdoped regime) may yield an underlying FS
whose volume is too large \cite{NaCCOC,Shen05}.

Another method used in extracting the Luttinger surface is the `maximal gradient method'.
The method is based on the fact that the
FS is given by the set of $k$-values for which the momentum
distribution function $n_k$ shows a jump discontinuity. When this
discontinuity is smeared out, say, by thermal broadening or a
small gap, the gradient of $n_k$, $|\nabla n_k|$, is assumed to be
maximal at the locus of the underlying FS.

We calculated $|\nabla n_k|$ within RMFT and show our results in
Fig.~2. We see that the maximal gradient surface is very sensitive
to the presence of even small gaps. For example, the
superconducting gap at $x=0.25$ is quite small. Nonetheless, the
electron-like Luttinger surface (determined by $\xi_k\equiv0$) is
not clearly revealed by the ridges in $|\nabla n_k|$. Similar
deviations of $|\nabla n_k|$ from the underlying surface are also
obtained from a high temperature expansion of the $t-J$
model \cite{Putikka98} and dynamical cluster approximation in
the Hubbard model \cite{Maier02}. We conclude that the
maximal gradient method alone cannot be used to determine the
underlying FS unambiguously from ARPES data.

The notion that the underlying FS of a pseudogapped or a
superconducting state is identical to the Luttinger surface is
only approximately correct \cite{Dzyaloshinskii03,rosch}. In the
Fermi liquid state of normal metals, the FS satisfies the
Luttinger sum rule; the volume enclosed by the FS is identical to
the total number of conducting electrons. But, in a
superconductor, the chemical potential is generally renormalized
and is a function of the superconducting order parameter,
$\mu=\mu_{SC}(\Delta)$. The number of states $n_{Lutt}(\Delta)$
enclosed by the resulting Luttinger surface, $\xi_k\equiv0$, then
deviates from the true particle number $n$, as the results in
Fig.~3 show. However, this effect is small (a few percent) and
unlikely to be discerned experimentally. The discrepancy between
$n_{Lutt}(\Delta)$ and $n$ vanishes when particle-hole symmetry is
present. Further, it changes sign when the geometry of the
Luttinger surface changes from hole-like to electron-like, as seen
in Fig.~3.

Finally, we focus on the influence of the strong electron-electron
interactions on the geometry of the Luttinger surface close to
half filling. The Cu-O planes of the HTSC are characterized by a
nearest neighbor (NN) hopping parameter $t\approx 300~\mbox{meV}$
and a next nearest neighbor (NNN) hopping parameter $t'\approx
-t/4$. These parameters are the bare parameters, and determine the
dispersion relation,
\begin{equation}
\epsilon_k\ =\ -2t\Big(\cos(k_x)+\cos(k_y)\Big)
-2t'\Big(\cos(k_x+k_y)+\cos(k_x-k_y)\Big)\ , \label{epsilon_k}
\end{equation}
in the absence of any electron-electron interaction.
On the other hand, true hopping processes are
influenced by the Coulomb interaction
$U\approx 12\,t$
leading to a renormalization of the effective
hopping matrix elements,
$$
 t\ \ \to\ \  \tilde{t}=\tilde{t}(U),\qquad
t'\ \ \to\ \  \tilde{t}=\tilde{t}'(U).
$$
Close to half filling we find $\tilde{t}\propto J=4t^2/U$ and
$\tilde{t}'\to0$, \textit{i.e.}, the frustrating NNN hopping is
renormalized to zero. This behavior is illustrated in Fig.~4. The
resulting Luttinger surface renormalizes to perfect nesting. A
similar behavior has been observed in recent variational studies
of organic charge transfer-salt superconductors \cite{Liu05}. At
half filling, the so-called Marshall sign rule is valid rigorously
in the absence of frustration. The degree of effective frustration
can then be estimated by the size of deviation from the Marshall
sign rule as a function of the (frustrating) bare $t'$. A
numerical study has found, that the Marshall
sign rule remains valid even for small but finite $t'$,
\textit{viz.}, the effective frustration renormalizes
to zero \cite{heisenberg}.
This behavior is in agreement with the results presented in
Fig.~4, and is unique to strong coupling theories such as RMFT.

We showed that the accurate determination of the underlying FS in
underdoped HTSC is a difficult task and that analysis of the
experimental data alone is often insufficient for an unambiguous
determination of the FS. Commonly used methods like the zero
frequency spectral intensity or the gradient of $n_k$ can yield
significant deviations from the true Luttinger surface as shown in
Fig.~1 and 2. Indeed, a clear distinction between electron- and
hole-like underlying FS cannot be made solely from analyses of
spectral intensity maps when the gaps are large. Such analyses
have to be supplemented by a minimal modelling of the gapped
states. Furthermore, the underlying FS in the pseudogapped or
superconducting state fulfills Luttinger theorem only
approximately, owing to the dependence of the chemical potential
on the superconducting gap. We also demonstrated that the strong
correlations renormalize the ratio $\tilde t'/\tilde t$ near half
filling, yielding a Luttinger surface which is perfectly nested.
This suggests in a very natural way that the strong coupling mean
field superconducting state is unstable to antiferromagnetism at
low doping. Our findings resulting from the combined
effects of strong correlations and $d$-wave superconductivity, allow for a more
precise interpretation of experiments that determine the FS of
HTSC.

\newpage

\begin{figure}[t]
 \includegraphics[height=0.43\textwidth,keepaspectratio]
   {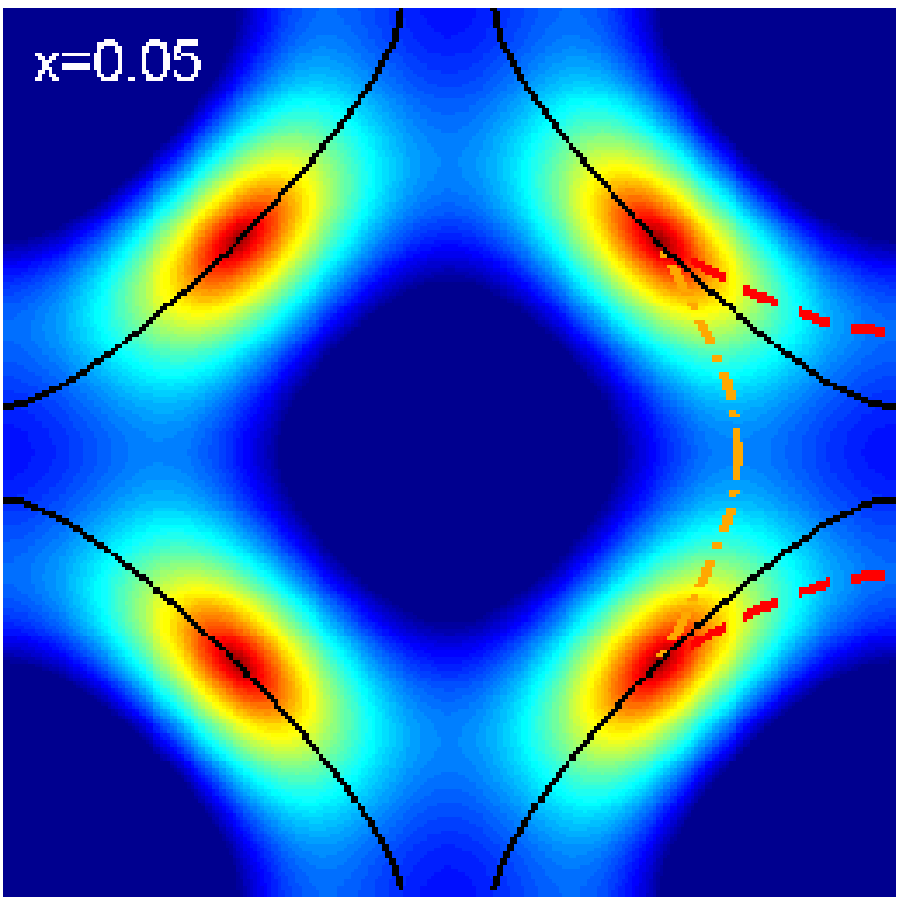}
 \includegraphics[height=0.43\textwidth,keepaspectratio]
   {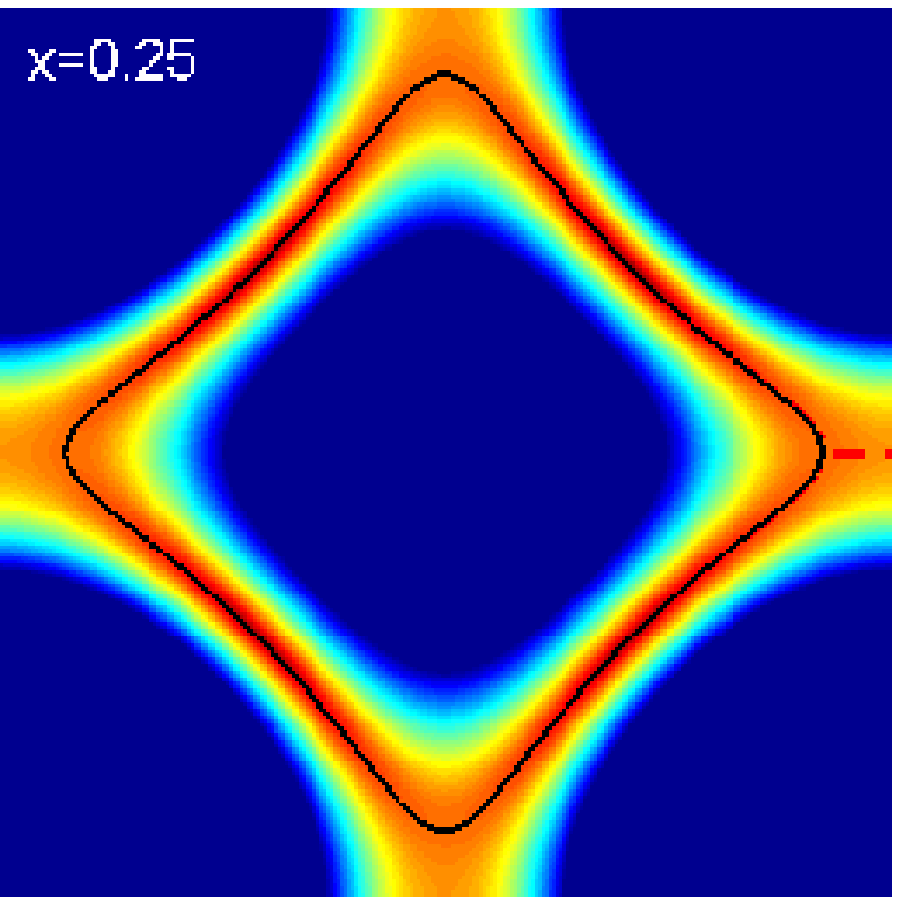}
\caption{The zero frequency spectral intensity (deduced from the
inverse of $E_k$) in the first Brillouin zone for hole dopings
$x=0.05$ (left) and $x=0.25$ (right). The color coding blue/red
corresponds to the low/high zero frequency spectral intensity. The
ridges of maximal intensity are indicated by the (dashed) red and
(dashed-doted) orange lines respectively, the Luttinger surface by
the black line.} \label{density_plotsEK}
\end{figure}

\begin{figure}[t]
 \includegraphics[height=0.43\textwidth,keepaspectratio]
  {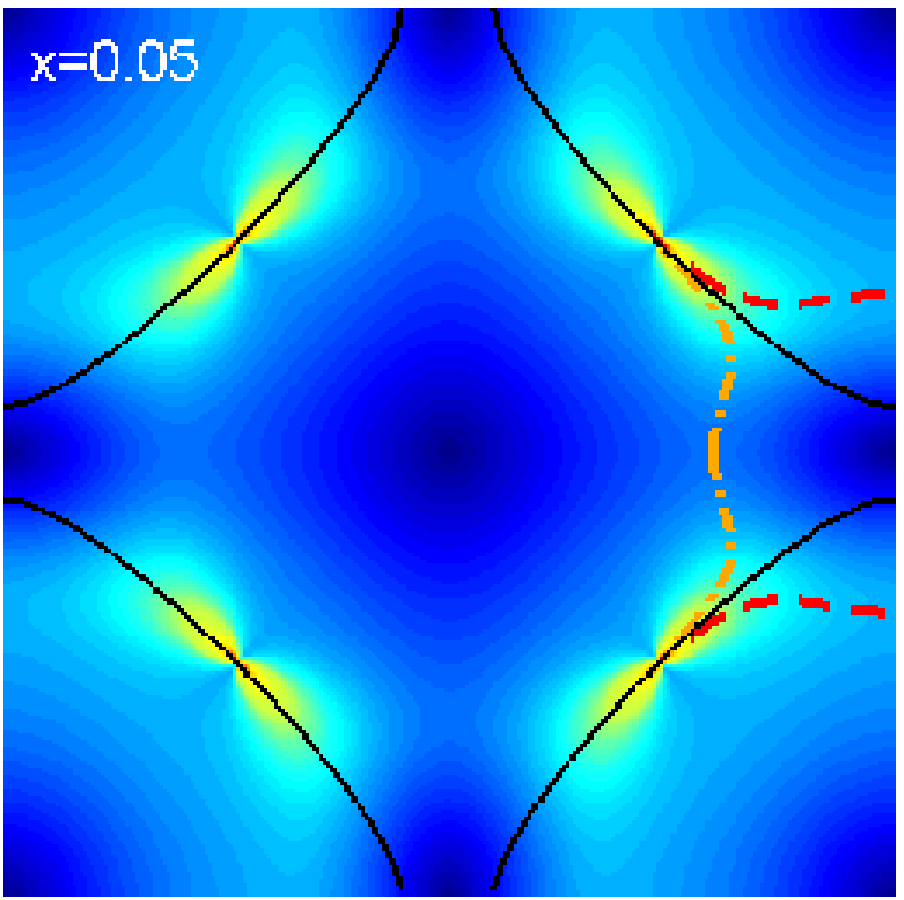}
 \includegraphics[height=0.43\textwidth,keepaspectratio]
  {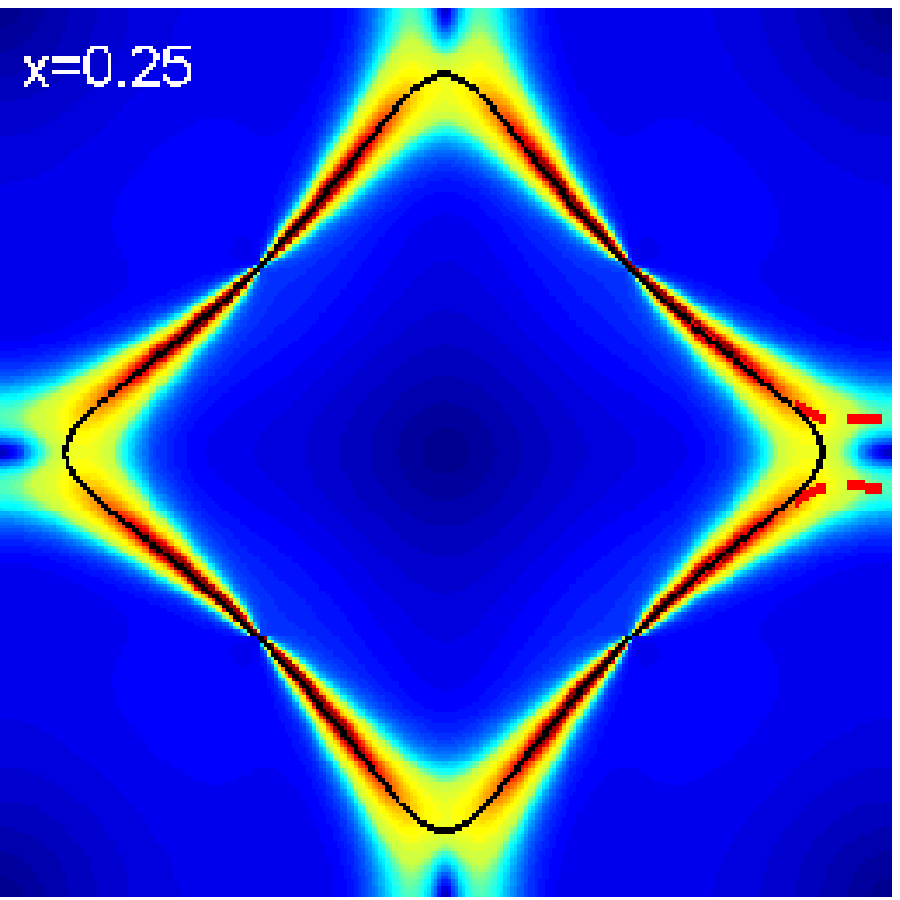}
\caption{The gradient of the momentum distribution function,
$|n_k|$, in the first Brillouin zone, for hole dopings $x=0.05$
(left) and $x=0.25$ (right). The color coding is blue/red for
small/large values of $|\nabla_k n_k|$. The ridges of maximal
$|\nabla_k n_k|$ are indicated by the (dashed) red and
(dashed-doted) orange lines respectively, the Luttinger surface by
the black line.} \label{density_plotsNK}
\end{figure}

\begin{figure}[t]
\centering
\includegraphics*[width=0.65\textwidth]{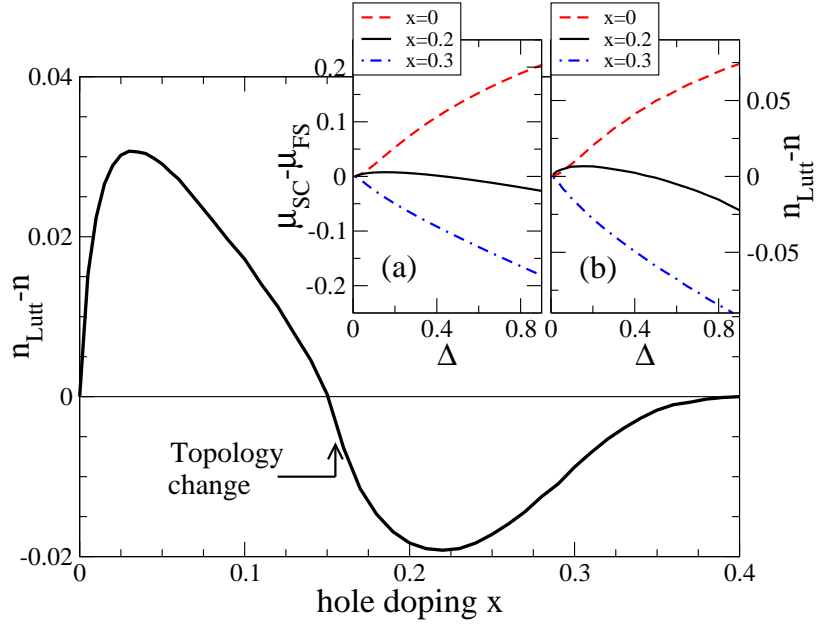}
\caption{The deviation, $n_{Lutt}-n$, of the actual volume
 of the Luttinger surface from the Luttinger sum-rule,
 as a function of hole-doping $x$.
 Calculations are performed by RMFT ($t'=-t/4$, $U=12t$).
 The deviation is minimal when the topology of the Luttinger
 surface changes from hole-like to electron-like.\newline
 Inserts (a) and (b): Model calculation for
 the renormalization of the chemical potential,
 $\mu_{SC}-\mu_{FS}$, and the resulting $n_{\rm Lutt}-n$
 as a function of the $d$-wave order
 $\Delta$, for various doping $x$. }
  \label{noLuttinger}
\end{figure}

\begin{figure}[t]
\centering
\includegraphics*[width=0.85\textwidth]{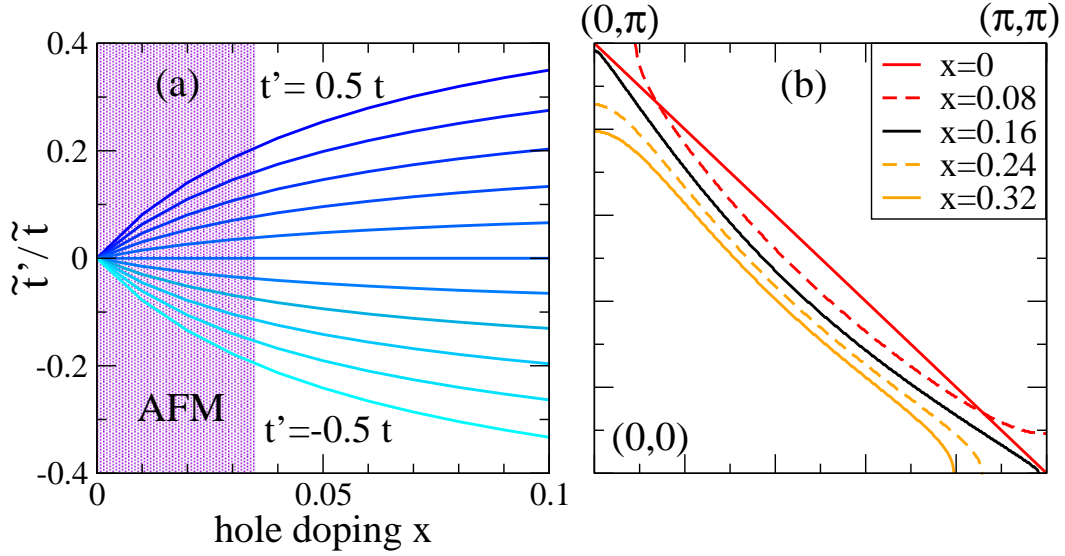}
\caption{(a) Renormalization of the next nearest neighbor
 hopping amplitude, $t'\to\tilde t'$, as a function
 of hole doping concentration $x$ for various values of bare $t'$.
 All effective $\tilde t'$ are renormalized to zero at half
 filling by the large Coulomb repulsion. We highlight the region
 for which we expect the superconducting $d$-wave state to become
 unstable against antiferromagnetism (AFM) due to the nearly
 perfect nesting of the Luttinger surface.
 \newline
 (b) The geometry of the Luttinger surface
 for the high temperature superconductors ($t'=-t/4$). The change is
 non-monotonic for small doping $x$, when the Luttinger surface is renormalized to
 perfect nesting due to the strong Coulomb interaction. For $x=0.16$,
 the topology of the Luttinger surface changes from hole-like to
 electron-like. \newline
 Calculations are performed for the Hubbard model with $U=12t$, using RMFT.
}
\label{nesting}
\end{figure}


\begin{thebibliography}{99}


\bibitem{Damascelli03} Damascelli, A., Hussain, Z., \& Shen, Z.-X.
 (2003) Rev. Mod. Phys. {\bf 75}, 473.


\bibitem{campuzano_rev} Campuzano, J.~C., Norman, M.~R., \&
Randeria, M. (2004) in \emph{Physics of Conventional and Unconventional
Superconductors}. (Springer); cond-mat/0209476.

\bibitem{nature_rev} For a recent overview,
see the special issue on high temperature
superconductivity in Nature Physics, {\bf 2}, (March 2006).


\bibitem{Pines_Nozieres} Pines, D. \& Nozi\'eres, P.
(1966) {\it The Theory of Quantum Liquids} (Addison-Wesley).

\bibitem{Loeser96} Loeser, A.~G., {\it et al.}
(1996) Science {\bf 273}, 325.


\bibitem{Marshall96} Marshall, D.~S., {\it et al.} (1996)
Phys Rev Lett. {\bf 76}, 4841.


\bibitem{Norman98} Norman, M.~R., {\it et al.} (1998)
Nature {\bf 392}, 157.


\bibitem{elstr} Edegger, B., Muthukumar, V.~N., Gros, C.,
\& Anderson, P.~W. (2006)
Phys. Rev. Lett. {\bf 96}, 207002.


\bibitem{Anderson87} Anderson, P.~W. (1987)
Science {\bf 235}, 1196.


\bibitem{Lee06} Lee, P.A., Nagaosa, N., \& Wen, X.-G. (2006)
Rev. Mod. Phys. {\bf 78}, 17.


\bibitem{Zhang88} Zhang, F.~C., Gros, C., Rice, T.~M., \& Shiba, H.
 (1988) Supercond. Sci. Tech. {\bf 1}, 36.


\bibitem{Dzyaloshinskii03} Dzyaloshinskii, I. (2003)
Phys. Rev. B {\bf 68}, 085113.


\bibitem{Gros94} Gros, C., Wenzel, W., Valent\'i, R.,
         H\"ulsenbeck, G., \& Stolze, J. (1994)
       Europhys. Lett. {\bf 27}, 299.

\bibitem{Mesot01}
Mesot, J., {\it et al.}
  (2001) Phys. Rev. B {\bf 63}, 224516.

\bibitem{yoshida05} Yoshida, T., {\it et al.}
, cond-mat/0510608.

\bibitem{NaCCOC} In particular, this effect is seen in
Ca$_{2-x}$Na$_x$CuO$_2$Cl$_2$ \cite{Shen05}, which also exhibits
quite a large pseudogap \cite{Kohsaka04}.


\bibitem{Shen05} Shen, K.~M., {\it et al.} (2005)
Science {\bf 307}, 901.


\bibitem{Putikka98} Putikka, W.~O., Luchini, M.~U., \& Singh, R.~R.~P.
 (1998) Phys. Rev. Lett. {\bf 81}, 2966.


\bibitem{Maier02} Maier, T.~A., Pruschke, T., \& Jarrell, M.
(2002) Phy.  Rev. B {\bf 66}, 075102.

\bibitem{rosch} Rosch, A.
 cond-mat/0602656.


\bibitem{Liu05} Liu, J., Schmalian, J., \& Trivedi, N.
(2005) Phys. Rev. Lett. {\bf 94}, 127003.


\bibitem{heisenberg} Richter, J., Ivanov, N.~B., \& Retzlaff, K.
 (1994) Europhys.  Lett. {\bf 25}, 545.

\bibitem{Kohsaka04} Kohsaka, Y., {\it et al.}
(2004) Phys. Rev. Lett.  {\bf 93}, 097004.

\end{thebibliography}
\end{document}